\documentclass[final,5p,times,twocolumn,authoryear]{elsarticle}

\usepackage[caption=false]{subfig}
\usepackage{graphicx}
\usepackage{placeins}
\usepackage{float}
\usepackage{calc}
\usepackage{amsmath,amssymb}
\usepackage{xcolor}
\usepackage{url}
\usepackage{hyperref}
\hypersetup{hidelinks}

\usepackage{aas_macros}
\usepackage[utf8]{inputenc}
\DeclareUnicodeCharacter{2243}{\ensuremath{\simeq}}
\DeclareUnicodeCharacter{2264}{\ensuremath{\leq}}
\DeclareUnicodeCharacter{2011}{-}
\DeclareUnicodeCharacter{2014}{---}
\DeclareUnicodeCharacter{2019}{\textquotesingle}

\newlength{\aasfigwidth}

\newenvironment{acknowledgments}{\section*{Acknowledgments}}{}

\newcommand{\be}{\begin{equation}}
\newcommand{\ee}{\end{equation}}
\newcommand{\JADEStwelve}{JADES-GS-z12-0~}
\newcommand{\JADESeleven}{JADES-GS-z11-0~}

\newcommand{\JADESzthirteen}{JADES-GS-z13-0~}
\newcommand{\JADESfz}{JADES-GS-z14-0~}
\newcommand{\JADESfo}{JADES-GS-z14-1~}
\newcommand{\unit}[1]{\mathrm{#1}}
\newcommand{\GeV}{\unit{GeV}}

\newcommand{\percc}{\unit{cm}^{-3}}

\newcommand{\Msun}{M_{\odot}}

\newcommand{\sigmav}{\langle\sigma v\rangle}

\journal{Astronomy \& Computing}

\begin{document}

\begin{frontmatter}

\title{Neural Network identification of Dark Star Candidates.~I. Photometry}

\author[colgate]{Sayed Shafaat Mahmud}
\author[brynmawr]{Adiba Amira Siddiqa}
\author[colgate]{Cosmin Ilie}

\affiliation[colgate]{organization={Colgate University},
            addressline={13 Oak Drive},
            city={Hamilton},
            postcode={NY 13346},
            country={USA}}

\affiliation[brynmawr]{organization={Bryn Mawr College},
            addressline={101 N Merion Ave},
            city={Bryn Mawr},
            postcode={PA 19010},
            country={USA}}

\begin{abstract} The formation of the first stars in the universe could be significantly impacted by the effects of Dark Matter (DM). Namely, if DM is in the form of Weakly Interacting Massive Particles (WIMPs), it could lead to the formation (at $z\sim 25-10$) of stars that are powered by DM annihilations alone, i.e. Dark Stars (DSs). Those objects can grow to become supermassive ($M\sim 10^6 \Msun$) and shine as bright as a galaxy ($L\sim 10^8 \Msun)$. Using a simple $\chi^2$ minimization, the first three DSs photometric candidates (i.e. \JADESeleven, \JADEStwelve, and \JADESzthirteen) were identified by \cite{Ilie:2023JADES}. Our goal is to develop tools to streamline the identification of such candidates within the rather large publicly available high redshift JWST data sets. We present here the key first step in achieving this goal: the development and implementation of a feed-forward neural network (FFNN) search for Dark Star candidates, using data from the JWST Advanced Deep Extragalactic Survey (JADES) photometric catalog. Our method reconfirms JADES-GS-z13 and JADES-GS-z11 as dark star candidates, based on the chi-squared goodness of fit test, yet it is $\sim10^4$ times faster than the Neadler-Mead $\chi^2$ minimization method used in \cite{Ilie:2023JADES}. We further identify six {\it new photometric} Dark Star candidates across redshifts $z \sim 9$ to $z \sim 14$. These findings underscore the power of neural networks in modeling non-linear relationships and efficiently analyzing large-scale photometric surveys, advancing the search for Dark Stars. 
\end{abstract}


\end{frontmatter}

\section{Introduction} \label{sec:intro}

The application of machine learning (ML) in astrophysics has grown rapidly over the past decade, transforming how researchers analyze large and complex datasets.  ML has proven to be a versatile and powerful tool used for uncovering patterns in high-dimensional data, allowing for key results ranging from classifying galaxies~\citep[e.g.][]{zeraatgari2024machine, aguilar2025morphological, vavilova2021machine, cheng2020optimizing}, identifying exoplanets~\citep[e.g.][]{garvin2024machine,yu2019identifying, de2022identifying, cantero2023sodinn, sahlmann2025machine}, detecting gravitational waves~\citep[e.g.][]{marx2025machine,george2018deep, cuoco2020enhancing, cuoco2025applications, mitra2023exploring}, to mapping the cosmic web~\citep[e.g.][]{hong2021revealing, rodriguez2018fast}. Among ML techniques, Feed Forward Neural Networks (FFNNs) have emerged as a cornerstone in astrophysics due to their ability to model highly non-linear relationships, enabling applications such as photometric redshift estimation~\citep[e.g.][]{daza2025pau,razim2021improving, curran2021qso, treyer2024cnn}, lensing mass reconstruction\citep[e.g.][]{gupta2021mass}, and supernova classification~\citep[e.g.][]{villar2020superraenn, bairouk2023astronomical}. 

As the volume of data from current powerful instruments such as the James Webb Space Telescope (JWST), Euclid or the Vera Rubin Observatory continues to grow, and in view of the upcoming launch of the Roman Space Telescope (RST), the integration of ML in astrophysical and astronomical pipelines becomes increasingly essential. Neural networks, in particular, have been used to tackle complex problems involving noisy and incomplete datasets, making them invaluable for high-redshift studies \citep{stivaktakis2019convolutional, rastegarnia2022deep}. This paradigm shift enables researchers to analyze phenomena such as the early formation of galaxies, quasars, and stars at unparalleled precision. 

In this work we will develop ML tools for the identification of photometric Dark Star (DS) candidates in the publicly available data taken with the NIRCam instrument onboard the James Webb Space Telescope (JWST). Hypothesized to form in the high-redshift universe ($z \gtrsim 10$), Dark Stars are powered by dark matter (DM) annihilation rather than nuclear fusion \citep{spolyar2008dark,freese2008stellar}. They form out of pristine Hydrogen and Helium clouds at the center of DM microhalos, at redshifts $\sim [30,10]$, when said clouds are sufficiently cool to start collapsing under their own weight. The very high DM densities in those environments, further enhanced by the increased baryonic densities during the collapse phase~\citep[e.g.][]{Freese:2008dmdens},  lead to significant amounts of heating from Dark Matter annihilations. This, along with the poor cooling mechanisms available for zero metallicity clouds, can lead to a halting of the gravitational collapse of the proto-stellar gas cloud, well before it reaches the temperatures and densities necessary to ignite hydrogen nuclear fusion. As such, a new type of star is formed: a Dark Star. While the amount of DM inside such an object is negligible when compared to its total mass (which is dominated by H and He), the high efficiency with which DM annihilation produces energy, which in turn is mostly trapped inside the star, leads to Dark Stars in fact being extremely bright, and very puffy. They can grow as massive as a million suns, and shine as bright as an entire galaxy~\citep{Freese:2010smds}. As such, they are detectable in current observatories, such as JWST~\citep{Ilie:2012}, or the upcoming Roman Space Telescope (RST)~\citep{Zhang:2022}. 

Dark Stars (DSs) are not only targets for direct searches, but their properties and cosmological abundance are already subject to a variety of observational constraints. In particular, recent JWST observations can constrain the contribution of lower-mass DSs to the unexpectedly abundant luminous sources observed at high redshift and thereby restrict combinations of DS masses, lifetimes, and formation efficiencies \citep{IoccoVisinelli2024,Lei2025}. Independent limits arise from the large energy release associated with dark-matter annihilation: the resulting high-energy emission can contribute to the diffuse cosmic-ray and $\gamma$-ray backgrounds, placing bounds on the abundance of annihilation-powered DSs and on the underlying dark-matter parameters \citep{Yuan2011}, while related multi-messenger signatures, including diffuse neutrino emission, provide complementary probes of populations of massive DSs and their remnants \citep{Schwemberger2025}. DS formation and evolution can also alter the thermal and ionization history of the intergalactic medium, allowing CMB, reionization, and global 21-cm observations to constrain the abundance and properties of DS populations and the dark-matter models that produce them \citep{Qin2024,Gondolo2022}. Finally, gravitational waves provide a qualitatively different population-level test: \citet{GhodlaIlie2025} showed that an early population of massive black-hole seeds produced by the collapse of supermassive Dark Stars can subsequently generate, through SMBH mergers, a nanohertz stochastic gravitational-wave background comparable to that measured by pulsar timing arrays, thereby allowing PTA observations to probe the abundance and mass distribution of the remnants of the SMDS population.

One of the main motivations for the study of Supermassive Dark Stars comes from recent astronomical data, which poses two significant challenges to traditional models of the formation of the first stars and their BH remnants. Namely, JWST reveals a large number of very compact, yet ultra bright galaxies forming in the first billion years after the Big Bang~\citep{GLASSz13, Maisies:2022, z16.CEERS93316:2022, z17.Schrodinger:2022,JADES:2022a,JADES:2022b,Labbe:2022,JWSTRedMonsters:2024}. Those early galaxies (sometimes called ``blue monsters'')  must have converted baryons into stars incredible rates, much higher than anything ever observed in the local universe~\citep{Boylan-Kolchin:2023}. Dark Stars, being much cooler than regular nuclear fusion powered stars, could, in principle accrete baryons indefinitely~\citep{Freese:2010smds}, i.e. easily match the observed efficiencies. As such Dark Stars offer a possible solution to this puzzle, as discussed in detail in \cite{Ilie:2023UHZ1,ilie2025spectroscopicsupermassivedarkstar}. Moreover, data from JWST, in conjunction with X-ray observatories such as Chandra, reveals that the Supermassive Black Holes (SMBHs) powering the most distant quasars ever observed are too big to have grown from regular nuclear powered stars, instead, most likely, requiring the existence very massive (i.e. $M\gtrsim 10^4\Msun$) Black Hole seeds in the first few hundred million years after the Big Bang. Such SMDSs coudld instead be seeded by Dark Stars~\citep[e.g.][]{Spolyar:2008dark,Freese:2010smds, Banik:2019} and/or Direct Collapse Black Holes~\citep[e.g.][]{Loeb:1994wv,Belgman:2006,Lodato:2006hw,Natarajan:2017,barrow:2018,Whalen:2020,Inayoshi:2020}. For a review on Dark Stars see \cite{Freese:2016dark}. 

The first photometric Dark Star candidates have been identified in \cite{Ilie:2023JADES}. Disambiguating between Supermassive Dark Stars (SMDSs) and early galaxies can be done most easily via a smoking gun signature specific to the spectra of SMDSs: an absorption at 1640 \AA (restframe) due to He~II. \cite{ilie2025spectroscopicsupermassivedarkstar} uses spectroscopic data for some of the most distant objects ever observed taken with the NIRSpec instrument onboard JWST and shows that \JADESeleven, \JADESzthirteen, \JADESfz, and \JADESfo are all consistent with a Dark Star interpretation, both in terms of spectra and morphology. Most intriguingly, a tentative ($S/N\sim 2$) 1640~{\AA} absoprtion feature is found for \JADESfz, the second most distant object ever observed (at the time of writing). Despite those early encouraging results, identifying Dark Stars in vast photometric surveys remains a significant challenge. Traditional approaches, such as $\chi^2$ minimization methods used in \cite{Ilie:2023JADES}, are computationally expensive and struggle to scale for large datasets. As a first step in scaling our Dark Star fitting pipeline to automatically analyze the vast amounts of publicly available relevant JWST data, in this work we develop ML algorithms which use the potential of neural networks to overcome challenges in terms of detection and classification in large datasets. Studies have demonstrated the use of NNs for photometric redshift estimation~\citep[e.g.][]{Firth:2003,Hoyle:2016,chunduri2023deep}, stellar classification~\citep[e.g.][]{abd2023multi}, and galaxy morphology analysis~\citep[e.g.][]{A_Naim_1995, ball2001morphological, E_J_Kim_2017}, showcasing their ability to handle high-dimensional, noisy data with remarkable efficiency. Building on this foundation, we apply a feed-forward neural network (FFNN) to identify and characterize Dark Star candidates based on photometric data from JWST. Our approach leverages the power of neural networks to model non-linear relationships between photometric features and the physical properties of Dark Stars, such as mass and redshift, while ensuring computational efficiency.

There are two different ways via which a Dark Star can access a sufficiently high DM density reservoir: adiabatic contraction (AC) and DM capture~\citep{freese2010supermassive}. In the case of the former, DM densities at the center of DM microhalos where DSs form are enhanced by the steepening of the gravitational profile due to the collapse of the pre-stellar molecular gas cloud, and subsequently by the growth of the Dark Star itself. This is a purely gravitational effect. Using adiabatic contraction (AC)~\citep{Blumenthal:1985,Freese:2008dmdens} one can show that the central DM densities can attain values as high as $10^{16}~\GeV\percc$.  For tri-axial (i.e. not perfectly spherically symmetric) DM halos, many DM particles are on chaotic or box orbits, the central DM can be replenished and the DM density could be kept high for millions (to billions) of years. If/when this adiabatically contracted DM reservoir is depleted, the star begins to collapse and becomes denser. If DM and baryons interact of each other, the process of DM capture becomes relevant. The star is now trappig DM from its environment, slowed down below the escape velocity via collisions with baryons inside it. Dark Stars powered by Captured DM are hotter than those powered via AC, and can attain temperature close to $5\times 10^4$~K~\citep{Freese:2010smds}. In view of this, the spectral energy distribution (SED) of SMDSs formed via AC and via DM capture are different.  As such, we train separate neural network models for these scenarios. Moreover, our NN models accommodate incomplete photometric datasets, an essential feature for analyzing real-world observations, which often suffer from missing photometric data due to instrumental or observational constraints.

\begin{figure}[!htbp]
    \centering
    \includegraphics[width=\linewidth]{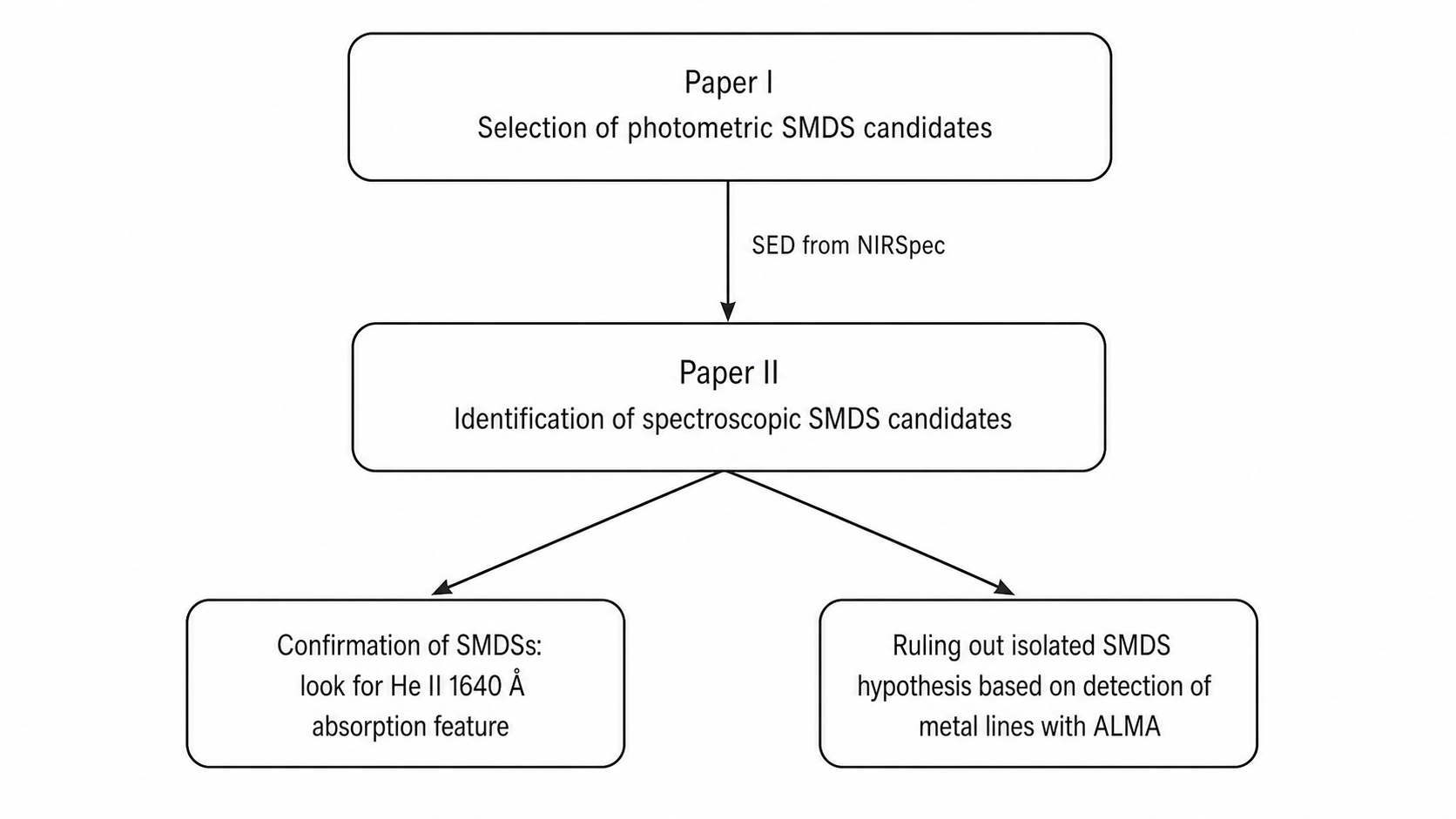}
    \caption{
        Schematic overview of the neural-network framework for identifying supermassive dark star (SMDS) candidates.
        In \emph{Paper~I} we develop a method to select photometric SMDS candidates.
        In \emph{Paper~II} we introduce a complementary method, applicable to objects with JWST/NIRSpec data, that uses the continuum spectral energy distribution to identify spectroscopic SMDS candidates.
        For these spectroscopic candidates, additional line-based follow-up is required, either to confirm the SMDS interpretation via a He~\textsc{ii}~$\lambda 1640$ absorption feature or to rule out the isolated SMDS hypothesis through the detection of metal lines with ALMA.
    }
    \label{fig:smds-nn-workflow}
\end{figure}

Through this work, we aim to address the dual challenges of precision and scalability in the search for Dark Star candidates using photometric JWST data taken with NIRCam. In the companion Paper~II~\citep{NNSMDSSpectra} we develop separate NN algorithms tailored for finding spectroscopic SMDSs candidates in the NIRSpec JWST data available for Lyman break high redshift objects. Disambiguation between the SMDSs candidates and high-z galactic counterparts cannot be made at the level of the continuum in the NIRSpec spectra. Specific emission or absorption features can be used to break this degeneracy. For example, the He~II 1640~{\AA} absorption feature is a smoking gun for SMDSs. Conversely, the identification of any metal lines via followup spectroscopy with ALMA would rule out the isolated SMDS interpretation for any of the candidates we identify via the methods presented in this work. The workflow discussed above is summarized in Fig.~\ref{fig:smds-nn-workflow}. 

By integrating machine learning techniques with astrophysical data, those two studies not only advance the understanding of Supermassive Dark Stars, but they also provide the cornerstone in establishing a framework for automatically analyzing the vast amount of publicly available JWST data in order to hunt for Dark Stars. The rest of the paper is organized as follows: Sec.~\ref{sec:Methods} details the neural network architecture and training procedure, in Sec.~\ref{sec: Results} we present our key findings, in Section~\ref{sec: Limitations} we discus the implications of our results and outline future directions, and end with conclusions in Sec.~\ref{sec: Conclusions}.

\section{Methods} \label{sec:Methods}
\subsection{Neural Network Architecture}
\begin{figure}[htbp]
    \centering
    \includegraphics[width=\linewidth]{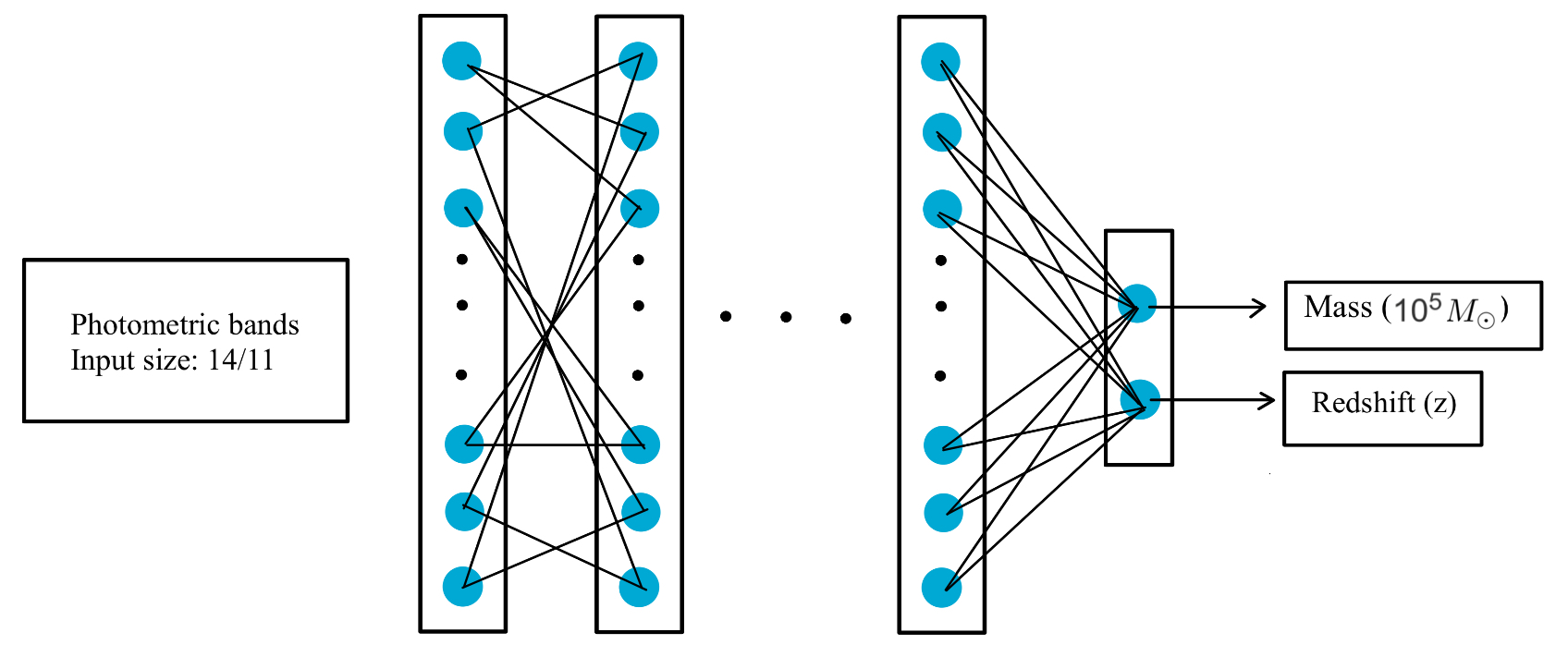}
    \caption{Schematic of the feed-forward neural network (FFNN) used in this work. The input consists of 14 or 11 photometric bands from JWST/NIRCam observations, depending on data availability. The network consists of fully connected hidden layers with ReLU activations and predicts two output parameters: stellar mass (in units of $10^5 M_\odot$) and redshift ($z$). Separate models are trained for adiabatic contraction and capture SMDSs formation scenarios.}
    \label{fig:ffnn_architecture}
\end{figure}

Feed-forward neural networks (FFNNs) are a class of models in machine learning, commonly used for regression and classification tasks \citep{rumelhart1986learning, lecun2015deep,goodfellow2016deep}. These networks consist of an input layer, one or more hidden layers, and an output layer, with information flowing in one direction through the network. Each neuron in one layer connects to every neuron in the next, allowing the model to represent complex relationships between input data and the quantities to be predicted. The model learns by adjusting the weights of these connections to minimize a loss function that measures the error between predicted and true values. We use gradient descent with the Adam optimizer \citep{kingma2015adam} to perform this optimization. Due to their flexibility and ability to capture non-linear patterns, FFNNs are particularly well-suited for analyzing astrophysical data, where such non-linearities are common \citep{bergmann2018deep, shallue2018identifying}.
In this work, we apply FFNNs to pre-selected photometric data from the medium and wide bands of NIRCam, targeting some of the most distant high-redshift sources identified by JWST surveys \citep{bunker2023jades, woodrum2023jades, haro2023spectroscopic}. Our goal is to fit this photometric data to theoretical models of Dark Stars, obtaining estimates for key physical parameters: redshift ($z$), stellar mass ($M$), and formation mechanism. We focus on the two hypothesized SMDSs formation scenarios—adiabatic contraction and DM capture—and design four separate neural networks accordingly. For each scenario, we develop two network configurations: one for complete datasets with 14 photometric bands and one for cases with only 11 bands. The 430M, 460M, and 480M filters are excluded in the latter case, as many high-redshift objects in the JADES catalog lack reliable flux measurements in these bands due to limited sensitivity.
Each network consists of fully connected layers with non-linear activation functions. For the full 14-band case, the input layer has 14 neurons, followed by three hidden layers: the first two with 128 neurons and the third with 64 neurons, all using ReLU activations \citep{goodfellow2016deep}. The output layer has two neurons corresponding to the predicted mass and redshift. For the networks using 11-band inputs, the architecture is kept the same except for the number of input neurons, allowing a direct comparison between results from full and partial photometric coverage.
To reflect the relative importance of the two distinct fit parameters: redshift ($z$) and SMDS mass ($M$), we use a weighted mean squared error (MSE) loss function:
 \begin{equation}\label{eq:Loss}
 \mathcal{L} = w_M \cdot \text{MSE}(M_{\text{pred}}, M_{\text{true}}) + w_z \cdot \text{MSE}(z_{\text{pred}}, z_{\text{true}}).
 \end{equation}

 The higher weight assigned to mass, $w_M = 2.0$ compared to $w_z = 1.0$, reflects the fact that stellar mass is the primary diagnostic for SMDS identification and is uniquely constrained by the photometric SED, whereas independent photometric redshift estimates are already available for the JADES catalog objects.
The goal of the neural network is to optimize this loss function. By optimizing this loss function we are essentially minimizing our error in predicting dark star mass and redshift. We train the networks using the Adam optimizer with a learning rate (i.e., the step size used to update model weights during training) of $10^{-3}$ for 300 epochs where each epoch corresponds to one full pass through the training dataset. A learning rate scheduler reduces the step size when improvements in validation loss stagnate, and early stopping is used to prevent overfitting. We adopt a batch size of 32 for the adiabatic contraction models and 128 for the capture models. The dataset is divided into training (72\%), validation (8\%), and test (20\%) subsets.
Before training, all input features and output targets are standardized using a common scaler to improve numerical stability and consistency. Model performance is evaluated on the test set using the $R^2$ score and mean absolute error (MAE), providing a quantitative measure of how well the networks recover redshift and mass values.

\subsection{Data Generation}
To construct our dataset, we simulate 10,000 photometric observations of Dark Stars spanning a range of stellar masses, redshifts, and formation mechanisms: either adiabatic contraction (AC) or DM capture. We simulate an equal number of AC and Capture cases. First, for each SMDSs mass, we compute the stellar atmospheric spectrum using the \texttt{TLUSTY} code \citep{hubeny2017TLUSTY}. For the composition of the stellar atmosphere we assume, as appropriate for Dark Stars, primordial BBN H/He ratios. The effective temperatures and surface gravity for each SMDSs modeled are taken from \citep{Freese:2010smds}, where SMDSs are modeled using the polytropic approximation. As shown in \citep{Rindler-Daller:2014uja}, using the 1D stellar evolution code \texttt{MESA}, SMDSs can be well modeled by $n=3$ polytropes. Throughout this work we assume, as done in most of the previous literature,  Dark Stars powered by 100 GeV WIMPs annihilating with $\sigmav=3\times 10^{-26}$cm$^3$s$^{-1}$, the canonical value that reproduces, via thermal production, the observed relic abundance of DM in the universe today.~\footnote{For a discussion about our restricted choice of $m_X$ and $\sigmav$, and our plans to overcome it in a future version of the NN algorithm presented here see Sec.~\ref{sec: Limitations}} We thus generate a large grid of \texttt{TLUSTY} high-resolution spectral fluxes as a function of wavelength, for a large variety of SMDSs masses. To simulate observations at cosmological distances, we shift the spectra to various redshifts ranging from 8 to 16, while accounting for the Lyman-alpha (Ly$\alpha$) break by excluding any flux contributions from wavelengths below the redshifted Ly$\alpha$ limit. This procedure mimics the absorption by the neutral H in the intergalactic medium, for objects at $z\gtrsim 6$~\citep{Gunn-Peterson:1965}.
We then convert these redshifted spectra into synthetic photometry by computing average fluxes within each of the JWST NIRCam filter bandpasses. This is done by integrating the stellar flux over the wavelength range of each filter, weighted by the filter’s throughput curve—i.e., its wavelength-dependent efficiency.
We use 14 NIRCam bands, covering both medium and wide filters, with throughput curves obtained from publicly available JWST documentation. 
To make the model more robust to noisy observations, we add Gaussian noise at the 10\% level to the fluxes in each band. This step helps produce more realistic inputs for our analysis. Each simulated data entry includes the 14 photometric band values, the corresponding redshift, the stellar mass, and the assumed formation channel (AC or capture). In total, we obtain 5000 photometric simulated data for AC case and 5000 photometric simulated data for Capture case. During training of each of the ML models, we perform a 72/8/20 split. This means 72\% of the data is used for training, 8\% is used for validation and 20\% of the data is used to test our model. For training our model we used the \texttt{Pytorch} \citep{paszke2019pytorch} library in python.

\subsection{Model Performance}

We evaluate the performance of our neural networks using training and validation loss plots, as well as predictions of mass and redshift compared to their true values. As we demonstrate below in this section, our model is very effective in accurately predicting dark star parameters under various input conditions. 

\begin{figure}[!htbp]
    \centering
    \begin{minipage}{0.49\linewidth}\centering
        \includegraphics[width=\linewidth]{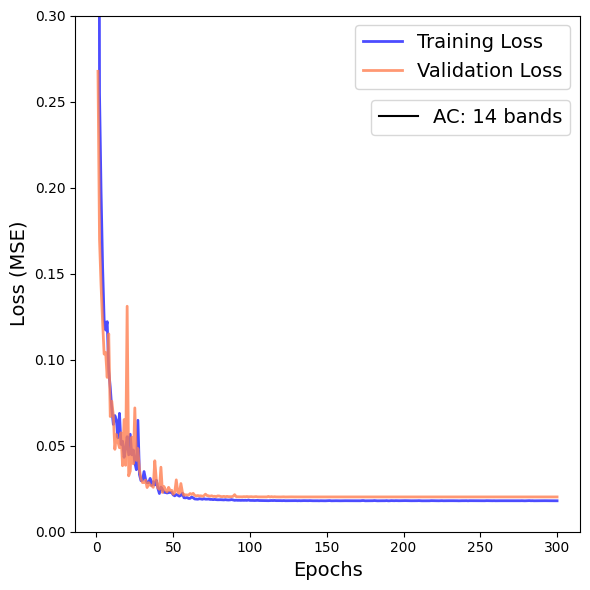}\\[-0.5ex]{\scriptsize (a)}
    \end{minipage}\hfill
    \begin{minipage}{0.49\linewidth}\centering
        \includegraphics[width=\linewidth]{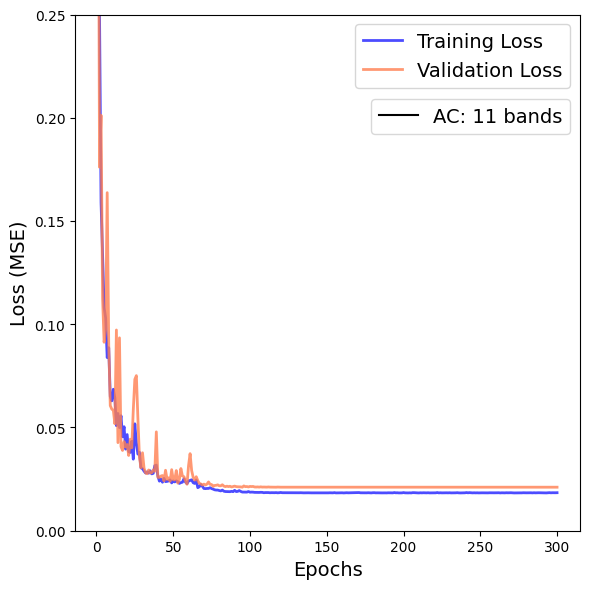}\\[-0.5ex]{\scriptsize (b)}
    \end{minipage}\\[0.4ex]
    \begin{minipage}{0.49\linewidth}\centering
        \includegraphics[width=\linewidth]{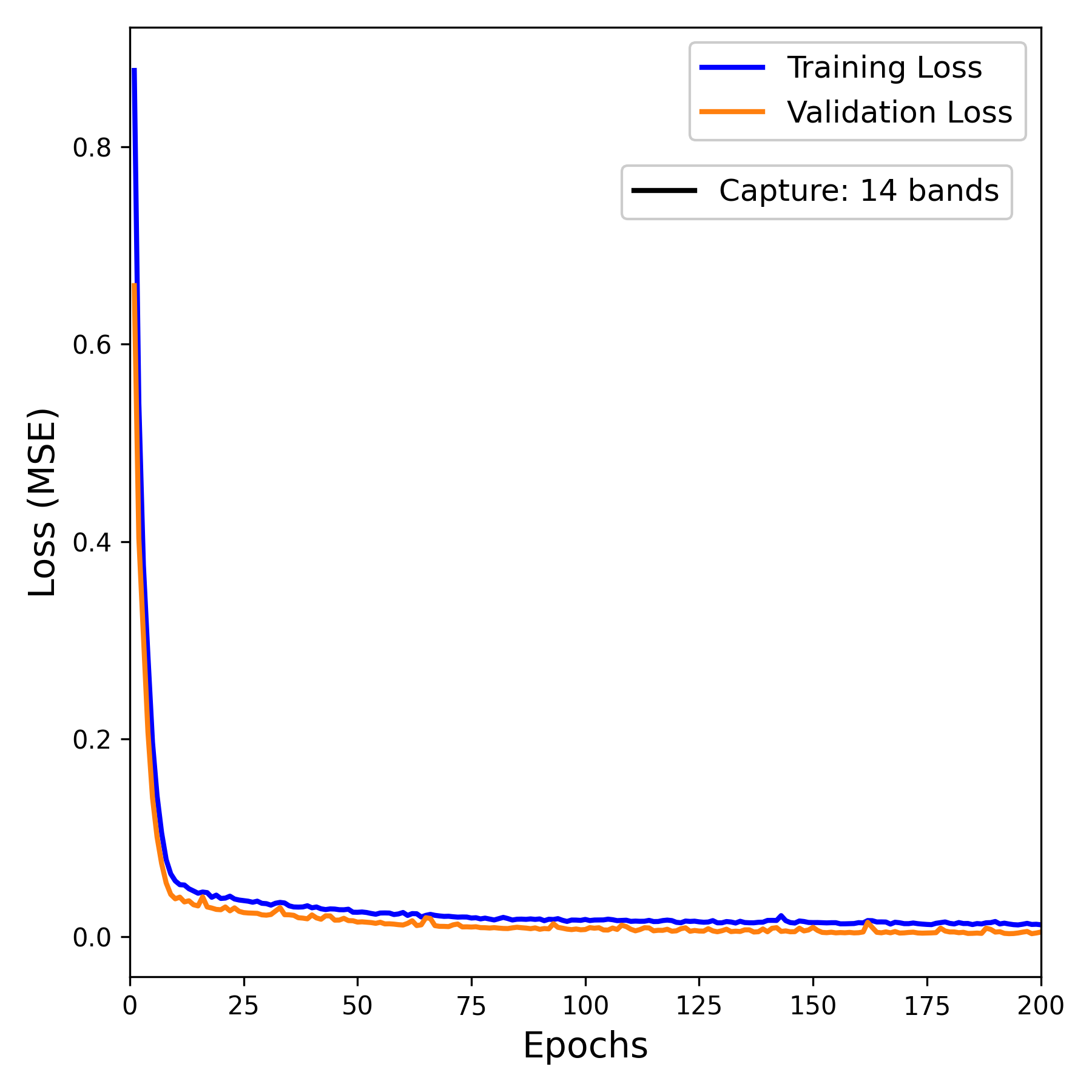}\\[-0.5ex]{\scriptsize (c)}
    \end{minipage}\hfill
    \begin{minipage}{0.49\linewidth}\centering
        \includegraphics[width=\linewidth]{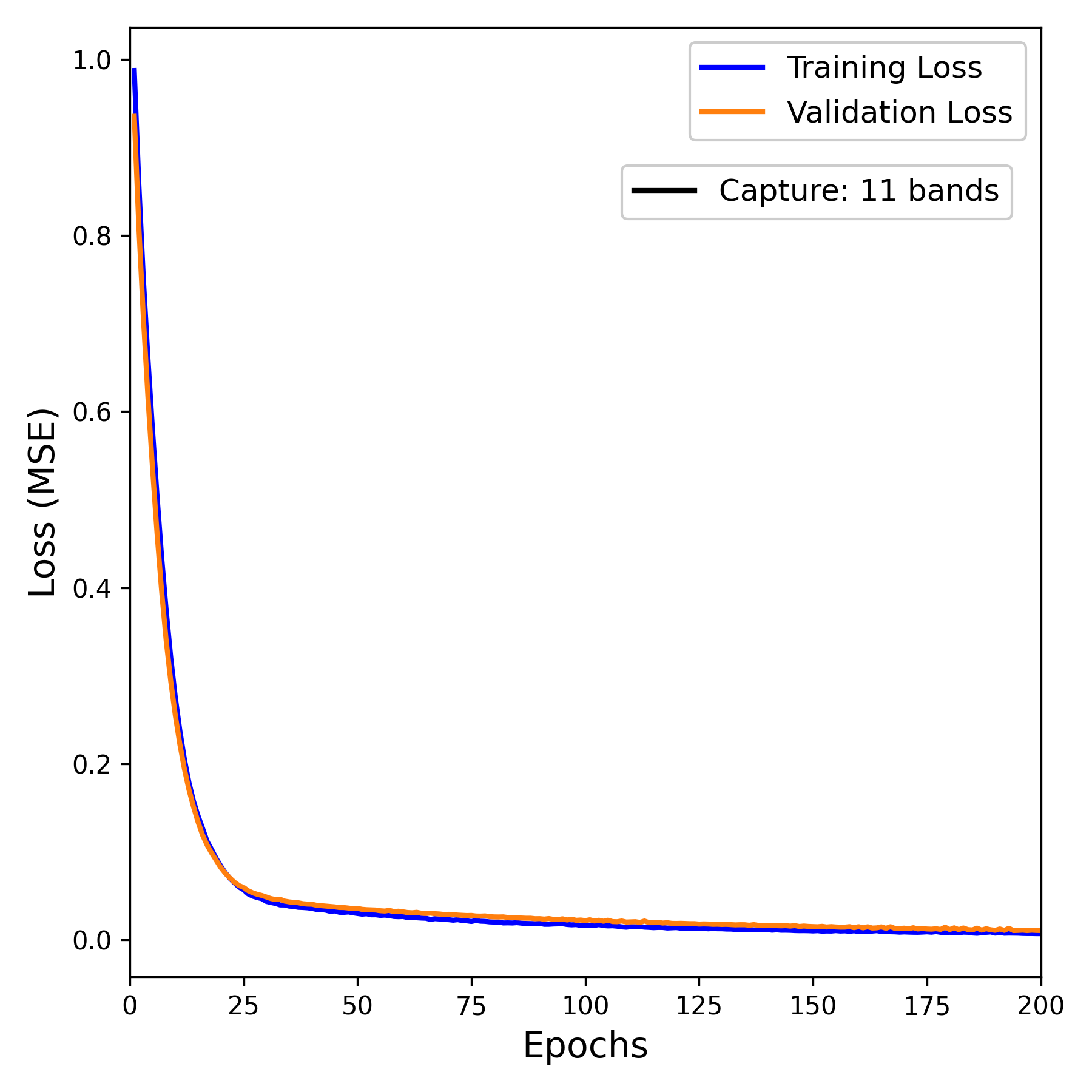}\\[-0.5ex]{\scriptsize (d)}
    \end{minipage}
    \caption{
        Training and validation loss, according to equation~\ref{eq:Loss}, for all four neural networks. 
        \textbf{Panel a:} Adiabatic contraction model with 14 inputs.
        \textbf{Panel b:} Adiabatic contraction model with 11 inputs. 
        \textbf{Panel c:} Capture case model with 14 inputs.
        \textbf{Panel d:} Capture case model with 11 inputs. 
        All models exhibit consistent convergence, with training and validation loss stabilizing after approximately 50 epochs. The small gap between training and validation loss indicates effective generalization to unseen data.
    }
    \label{fig:train-val-loss}
\end{figure}

\paragraph{Training and Validation Loss}
In order to assess how well the neural networks train, we monitor the decrease of loss (Equation~\ref{eq:Loss}) across training and validation dataset as a function of epochs. This gives us a sense of how close the model is predicting parameters close to the actual value as we iterate through the dataset. Figure~\ref{fig:train-val-loss} shows the training and validation loss for all four neural networks. These include the adiabatic contraction (AC) models with 14 and 11 inputs and the capture case models with the same configurations. For all networks, both training and validation loss converge smoothly, with no signs of overfitting. The loss stabilizes at a minimum value, demonstrating effective learning and generalization across the dataset.

\begin{figure}[!t]
\centering
\begin{minipage}{0.47\linewidth}\centering
\includegraphics[width=0.95\linewidth]{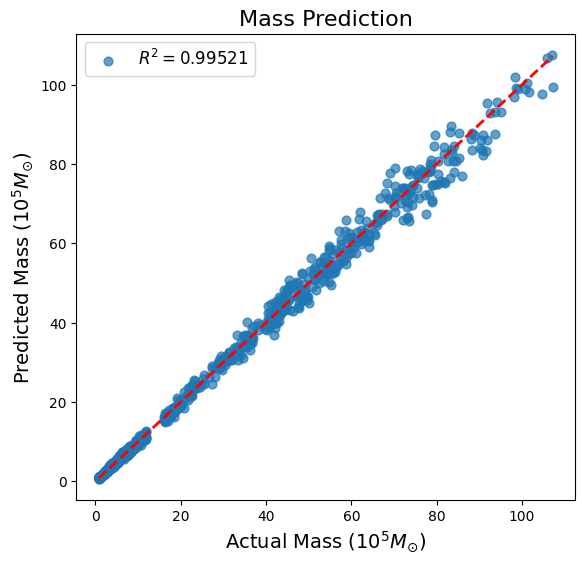}\\[-0.5ex]{\scriptsize (a) 14-input mass}
\end{minipage}\hfill
\begin{minipage}{0.48\linewidth}\centering
\includegraphics[width=0.95\linewidth]{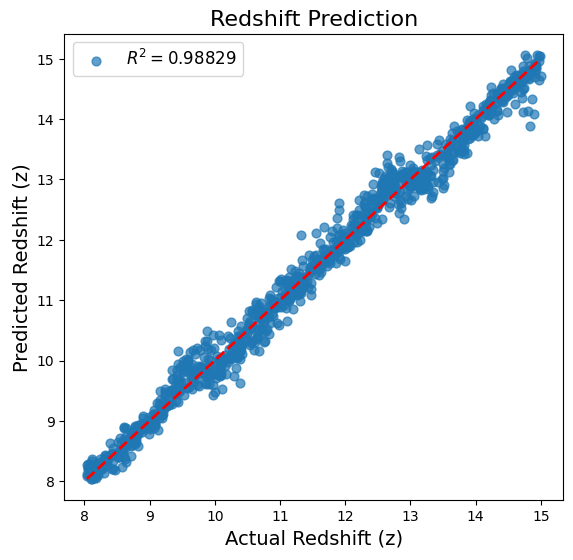}\\[-0.5ex]{\scriptsize (b) 14-input redshift}
\end{minipage}\\[0.2ex]
\begin{minipage}{0.48\linewidth}\centering
\includegraphics[width=0.95\linewidth]{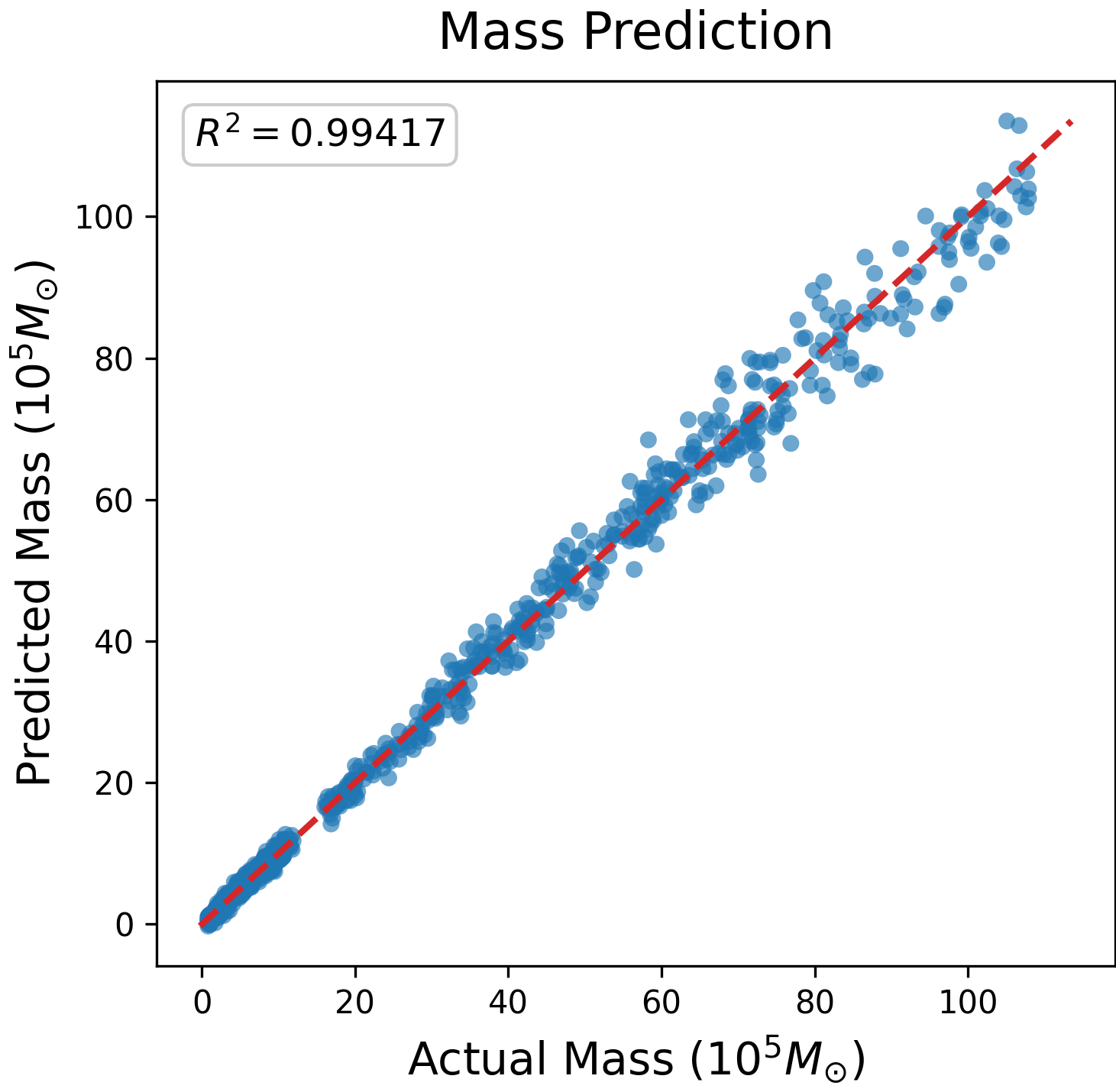}\\[-0.5ex]{\scriptsize (c) 11-input mass}
\end{minipage}\hfill
\begin{minipage}{0.48\linewidth}\centering
\includegraphics[width=0.95\linewidth]{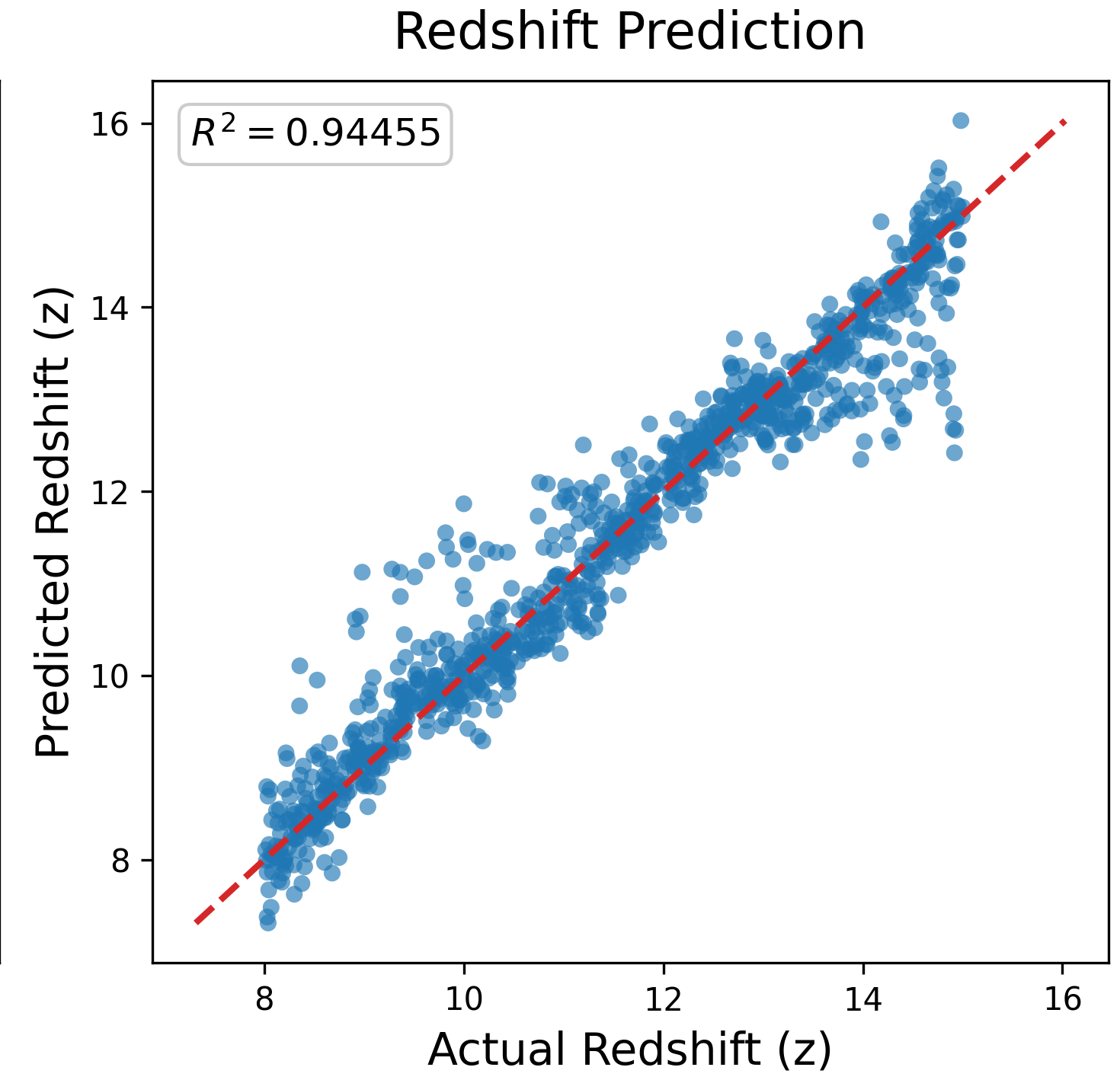}\\[-0.5ex]{\scriptsize (d) 11-input redshift}
\end{minipage}
\caption{Predicted versus true values of stellar mass and photometric redshift for the adiabatic contraction SMDSs models. 
        \textbf{Panels a--b:} Model with 14 inputs, achieving $R^2 = 0.995$ for mass and $R^2 = 0.989$ for redshift.
        \textbf{Panels c--d:} Model with 11 inputs, achieving $R^2 = 0.994$ for mass and $R^2 = 0.945$ for redshift. 
        The close alignment of points along the $y=x$ line reflects the high accuracy of the models in predicting both mass and redshift.}
 \label{fig:ac-pred-vs-true}
\end{figure}

\paragraph{Predicted vs. True Values: Adiabatic Contraction.}
Figure~\ref{fig:ac-pred-vs-true} illustrates the predicted versus true values for mass and redshift in the adiabatic contraction models. For the 14-input model (panel a), the predictions align closely with the true values, achieving $R^2 = 0.995$ for mass and $R^2 = 0.989$ for redshift. Similarly, the 11-input model (panel b) performs well, achieving $R^2 = 0.994$ for mass and $R^2 = 0.945$ for redshift. The results demonstrate that missing photometric bands slightly reduce accuracy but do not significantly impact overall model performance.

\begin{figure}[!t]
    \centering
    \begin{minipage}{0.47\linewidth}\centering
        \includegraphics[width=0.95\linewidth]{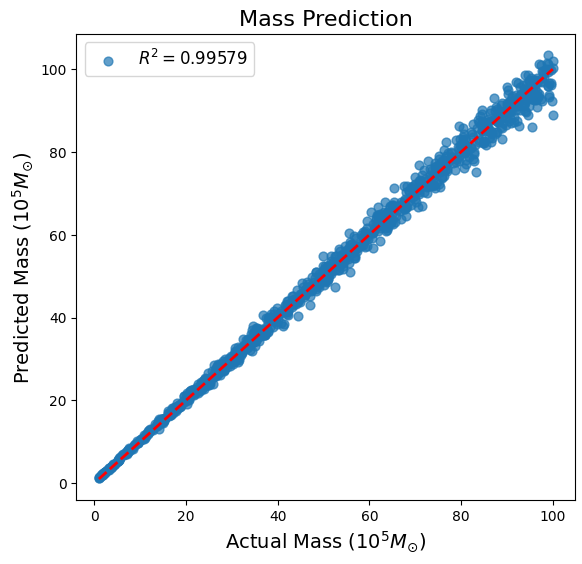}\\[-0.5ex]{\scriptsize (a) 14-input mass}
    \end{minipage}\hfill
    \begin{minipage}{0.48\linewidth}\centering
        \includegraphics[width=0.95\linewidth]{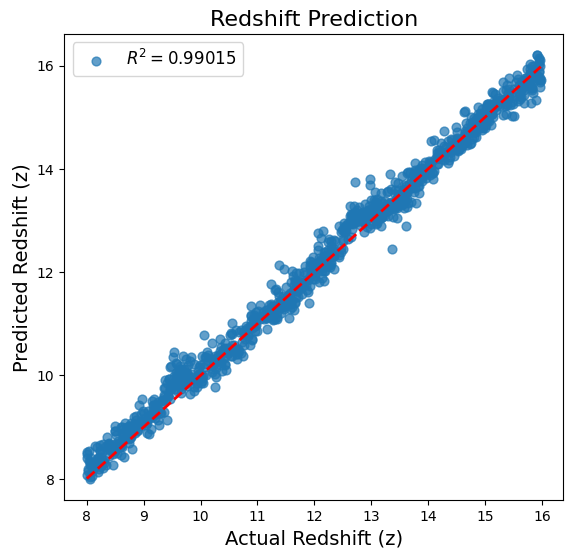}\\[-0.5ex]{\scriptsize (b) 14-input redshift}
    \end{minipage}\\[0.2ex]
    \begin{minipage}{0.48\linewidth}\centering
        \includegraphics[width=0.95\linewidth]{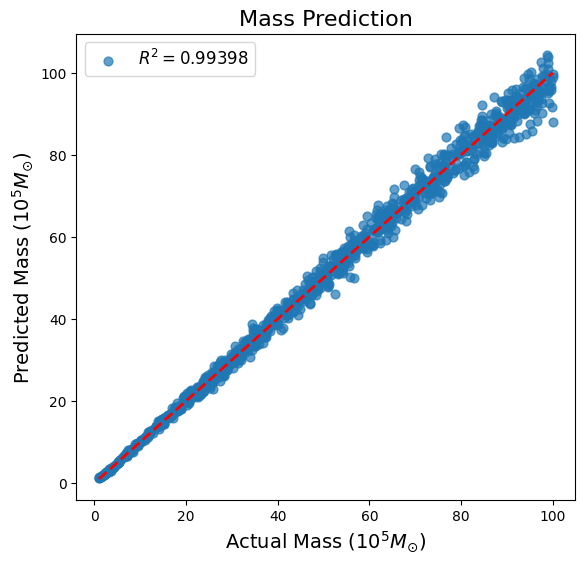}\\[-0.5ex]{\scriptsize (c) 11-input mass}
    \end{minipage}\hfill
    \begin{minipage}{0.48\linewidth}\centering
        \includegraphics[width=0.95\linewidth]{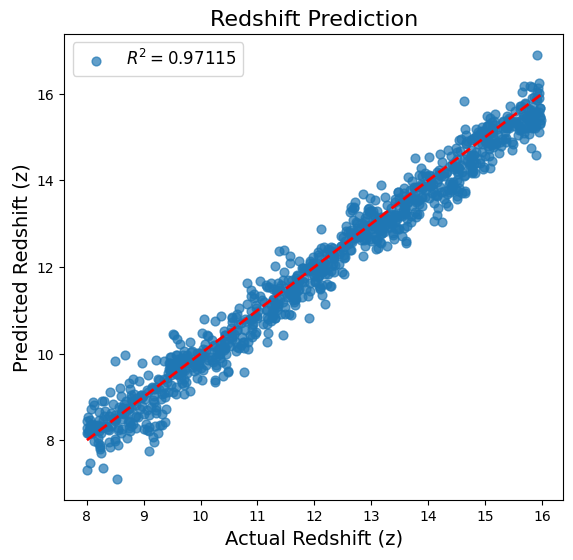}\\[-0.5ex]{\scriptsize (d) 11-input redshift}
    \end{minipage}
    \caption{
       Predicted versus true values of stellar mass and photometric redshift for the DM Capture SMDSs models. 
        \textbf{Panels a--b:} Model with 14 inputs, achieving $R^2 = 0.996$ for mass and $R^2 = 0.989$ for redshift.
        \textbf{Panels c--d:} Model with 11 inputs, achieving $R^2 = 0.993$ for mass and $R^2 = 0.963$ for redshift.
    }
    \label{fig:capture-pred-vs-true}
\end{figure}

\paragraph{Predicted vs. True Values: Capture Case.}
Figure~\ref{fig:capture-pred-vs-true} presents the predictions for the capture case models. Similar to the adiabatic contraction results, the 14-input model (panel a) achieves $R^2 = 0.995$ for mass and $R^2 = 0.989$ for redshift, while the 11-input model (panel b) achieves $R^2 = 0.993$ for mass and $R^2 = 0.971$ for redshift. Although the 11-input model shows a slight decrease in accuracy for redshift, the overall performance remains robust, validating the ability of the models to handle incomplete photometric data.

Our results presented in Figs.~\ref{fig:ac-pred-vs-true}-\ref{fig:capture-pred-vs-true} confirm the robustness of the models in predicting mass and redshift even when data in some photometric bands are missing. Having established that our FFNN model was trained successfully, we next proceed to apply it to actual data, as described in the next section. 

\FloatBarrier
\section{Results} \label{sec: Results}

\subsection{Confirmation of Previous Candidates}

\begin{figure}[!htbp]
    \centering
    \includegraphics[width=0.49\linewidth]{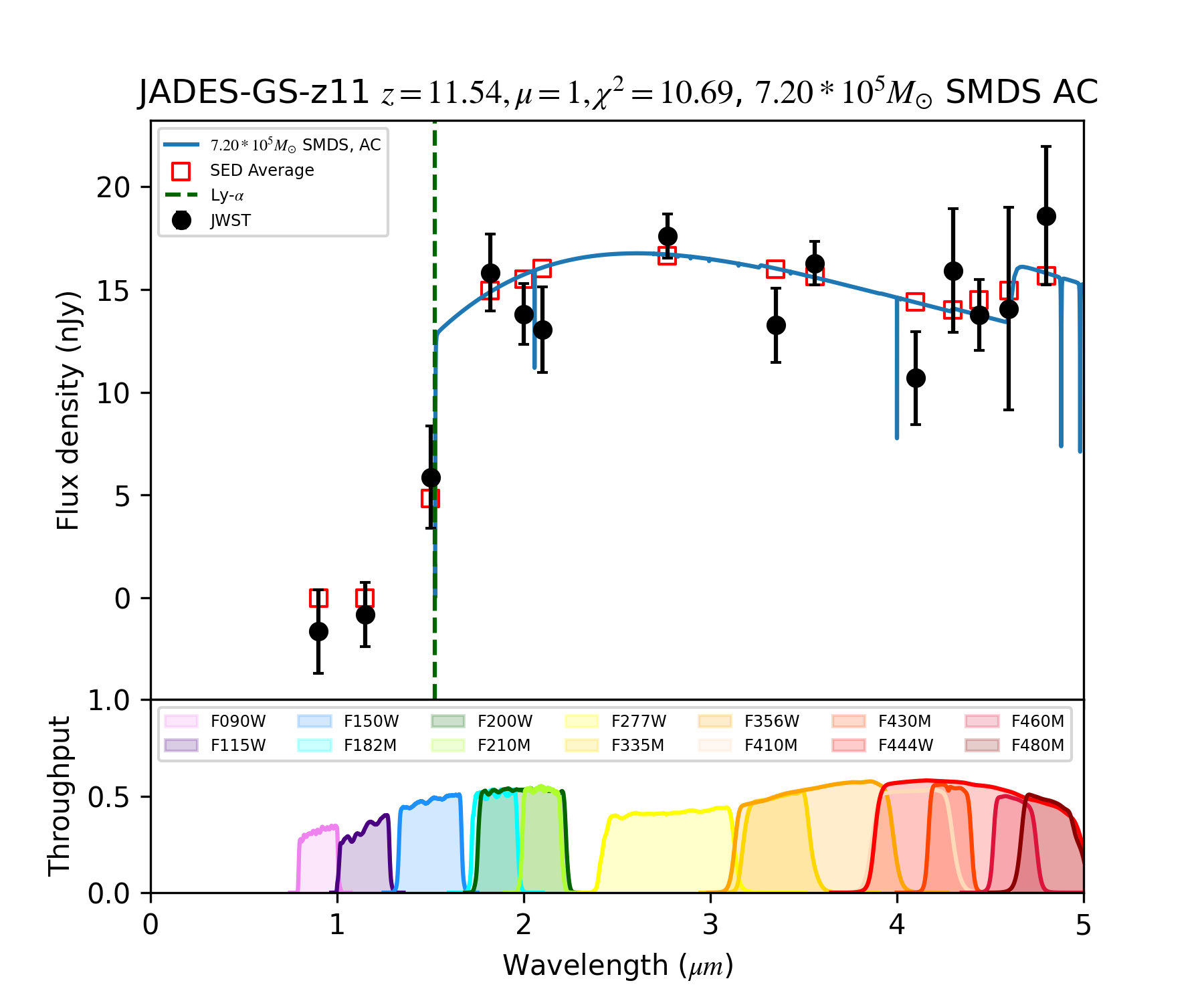}
    \includegraphics[width=0.49\linewidth]{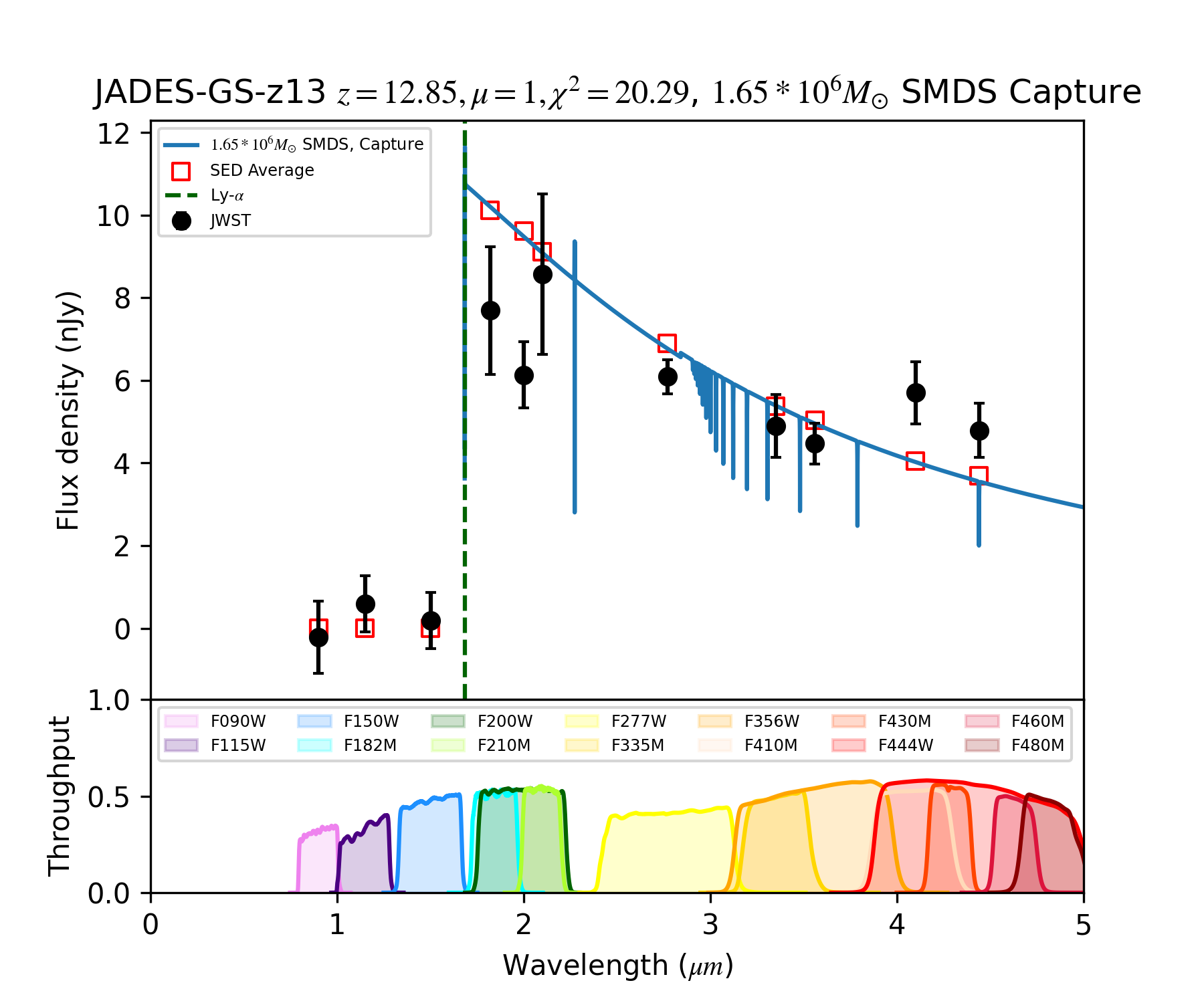}
    \caption{
        Comparison of predicted photometric flux density distributions to NIRCam data for JADES-GS-z11 and JADES-GS-z13, SMDSs photometric candidates first identified in \cite{Ilie:2023JADES}. The solid blue lines represent the FFNN predicted SEDs, with the red boxes representing average flux density in each photometric filter considered. The black points with error bars indicate JWST observations.
        \textbf{Left:} JADES-GS-z11 with $z=11.54$, $\mu=1$, $\chi^2=10.69$, and predicted stellar mass $7.20 \times 10^5 M_\odot$ using the adiabatic contraction (AC) model.
        \textbf{Right:} JADES-GS-z13 with $z=12.85$, $\mu=1$, $\chi^2=20.29$, and predicted stellar mass $1.65 \times 10^6 M_\odot$ using the capture case model.
         The throughput curves of the photometric bands are shown below for reference, demonstrating the wavelength range contributing to each band.
    }
    \label{fig:jades-z11-z13}
\end{figure}

The first three photometric SMDSs candidates were identified by \cite{Ilie:2023JADES}. Two of those (\JADESeleven and \JADESzthirteen) were subsequently confirmed to be consistent with a Dark Star interpretation even when deep spectroscopic NIRSpec data was used~\citep{ilie2025spectroscopicsupermassivedarkstar}. As such, we begin the application of our our FFNN to real data with \JADESeleven and \JADESzthirteen. 
In Figure~\ref{fig:jades-z11-z13} we present the flux density photometric distributions obtained for those two objects using our neural network-based approach. For \JADESeleven, we find a stellar mass of $7.20 \times 10^5 M_\odot$ at a redshift of $z=11.54$, with a $\chi^2$ value of $10.69$. Since for this object there is data in 14 NIRCam bands, $\chi_{crit}=22.4$, for the commonly assumed $\alpha=0.05$ significance level. Therefore according the the chi-squared statistic test, we can validate the null-hypothesis (in this case a Dark Star interpretation) for fit we find with the FFNN.  Similarly, for \JADESzthirteen, our method predicts a stellar mass of $1.65 \times 10^6 M_\odot$ at $z=12.85$, with a $\chi^2$ value of $20.29$. We note here that this $\chi^2$ value is slightly larger than 18.3, the critical value implied by the standard 0.05 significance level when we have data in 11 NIRCam bands. We find that for this particular fit we need to lower our significance level to 0.025 when using the chi-squared goodness of fit test. This might seem as a drawback of our FFNN, as for some objects, the fits it finds are slightly worse than those we identified in the past with the traditional Neadler-Mead algorithms in \cite{Ilie:2023JADES}. This can be traced back to our usage of a loss function based on mean square error rather than $\chi^2$ (see Eq.~\ref{eq:Loss}), as commonly done for FFNNs. However, the extreme increase in computational efficiency with which the FFNN identifies photometric candidates vastly compensates for this slight inaccuracy in the prediction of best fit parameters for some of the fits it might find. 

Note that \cite{Ilie:2023JADES} used a rather course stellar mass grid in order to generate \texttt{TLUSTY} templates for SMDSs. In order to compensate for this coarseness (i.e. three models per decade in mass), a gravitational lensing boost factor ($\mu$) was used as a third free parameter, in addition to redshift ($z$) and SMDS mass ($M$). This was a reasonable choice, especially for SDSMs modeled via DM Capture, as, in this case, $\mu$ and $M$ are completely degenerate parameters. For models via AC this degeneracy is milder, yet it still exists. In this work we generate much finer \texttt{TLUSTY} grids for SMDSs, and, as such, pin $\mu=1$, unless otherwise determined by independent observations for any specific object.

\begin{table*}[!tbp]
    \centering
    \caption{Comparison of Neural Network Predictions (NN) and those made via Nedler-Mead (N-M) approach in \cite{Ilie:2023JADES}.}
    \label{tab:validation}
    \begin{tabular}{lcccc}
        \hline
        \textbf{Object Name} & \textbf{$\mu\times M_{SMDS}$ ($10^5 M_\odot$)} (NN) & $\mu\times$\textbf{$M_{SMDS}$($10^5 M_\odot$)} (N-M) &\textbf{$z_{phot}$} (NN) & \textbf{$z_{phot}$} (N-M) \\
        \hline
        JADES-GS-z11   & $7.2$ & $7.5$ &11.54 & 11.66\\
        JADES-GS-z13   & $16.5$ & $15$ &12.85 & 13.9\\
        \hline
    \end{tabular}
\end{table*}

In summary,  the FFNN we designed and implemented in this work, and the traditional method (Neadler-Mead) we used in \cite{Ilie:2023JADES} lead to similar findings in terms of parameters for best fit SMDSs models to photometric data (see Table~\ref{tab:validation}). The neural network approach offers a significant advantage in computational efficiency, operating approximately $10^4$ times faster than the Nelder-Mead algorithm. This acceleration is primarily due to the parallelized nature of neural network inference, as opposed to the iterative optimization employed in Nelder-Mead. As alluded to before, another improvement with respect to the analysis pipeline used in \cite{Ilie:2023JADES} is the much finer SMDS mass grid used here, allowing us to capture subtleties in the spectral energy distribution (SED) that might otherwise be missed.

\subsection{Detection of New Candidates}

We next apply our neural network model to several other objects observed by the JADES survey \citep{hainline2024cosmos}. We pre-select them based on the following conditions described below. First,  they should be consistent with Lyman break objects and with photometric redshifts previously reported to be greater than nine, as SMDSs at lower redshifts are expected to be very rare, as shown by \cite{Ilie:2012}. Moreover, we select objects for which observations are done in at least 11 NIRCam photometric filters, in order to increase the constraining power.
We report here the identification of six additional SMDSs photometric candidates. Three of those are modeled by SMDSs formed via the adiabatic contraction (AC) mechanism, whereas the other three are powered by captured DM. This demonstrates the versatility of our approach in handling diverse datasets and physical scenarios. The neural network predicts mass and redshift that effectively gives the observed spectral energy distributions (SEDs) across photometric bands ensuring robust predictions. Importantly, the $\chi^2$ values for all candidates fall within the 95\% confidence interval, affirming the statistical reliability of our detections.

\begin{figure*}[!tbp]
\centering
\begin{minipage}{0.32\textwidth}\centering
\includegraphics[width=\linewidth]{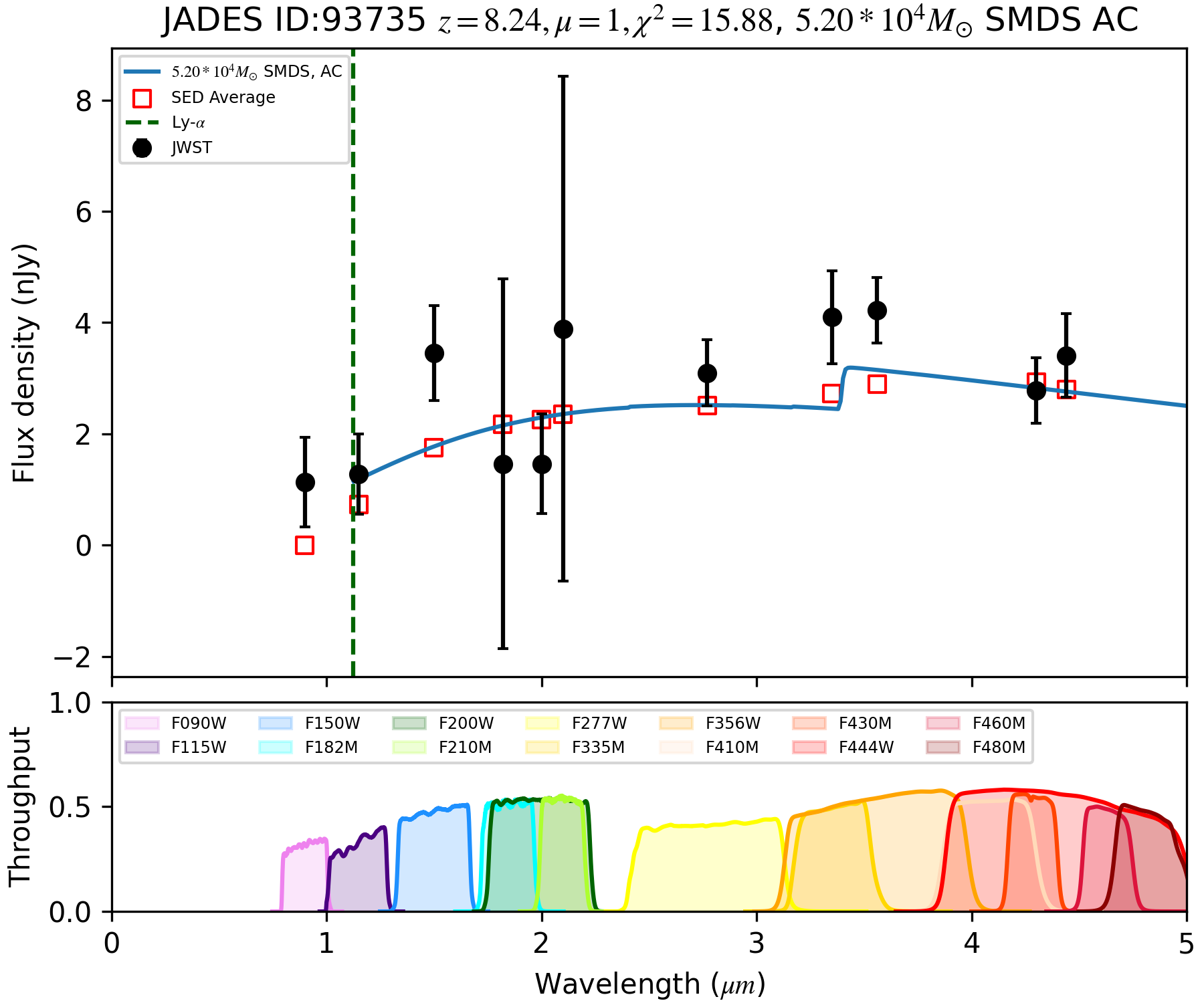}\\[-0.5ex]{\scriptsize (a)}
\end{minipage}\hfill
\begin{minipage}{0.32\textwidth}\centering
\includegraphics[width=\linewidth]{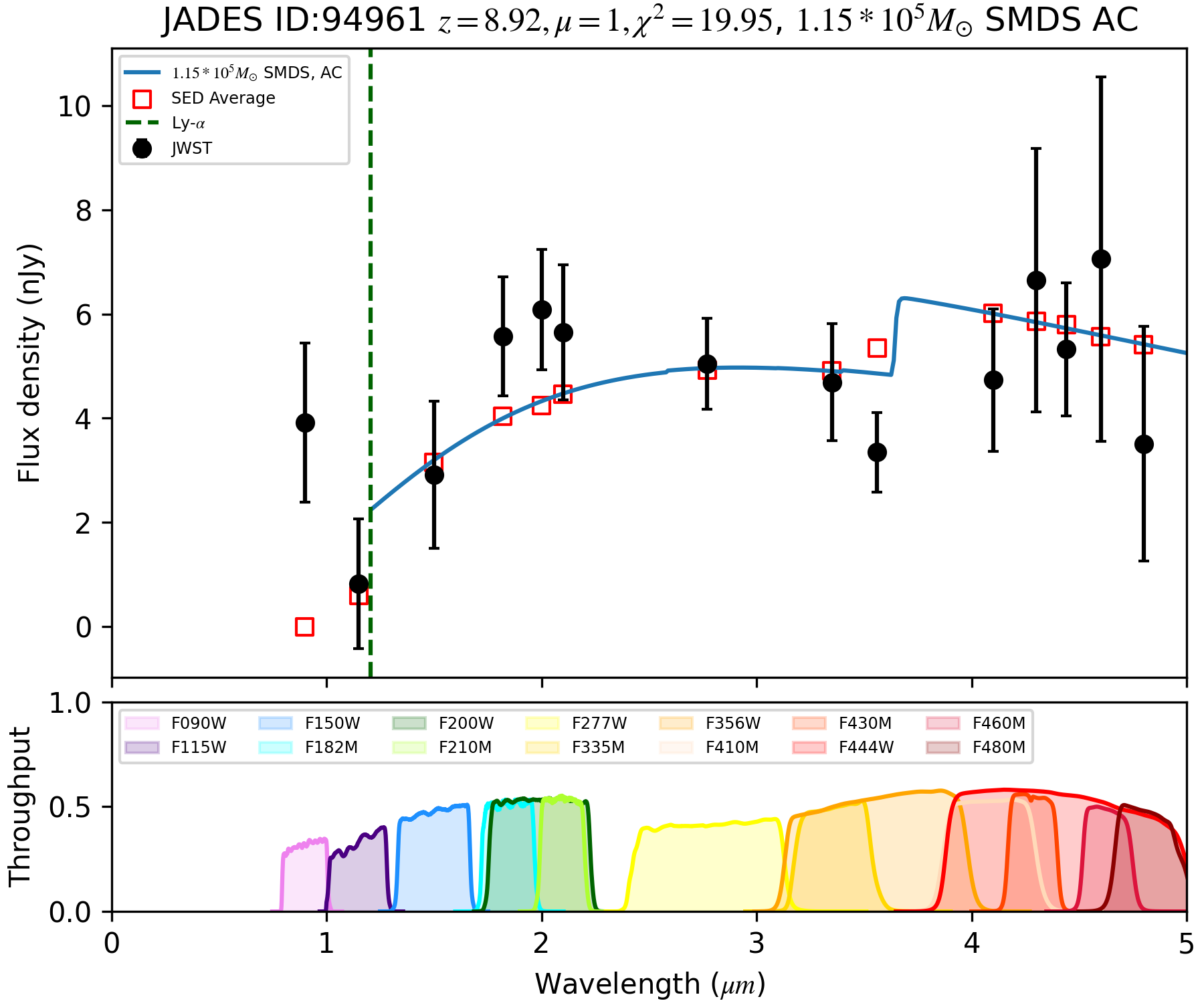}\\[-0.5ex]{\scriptsize (b)}
\end{minipage}\hfill
\begin{minipage}{0.32\textwidth}\centering
\includegraphics[width=\linewidth]{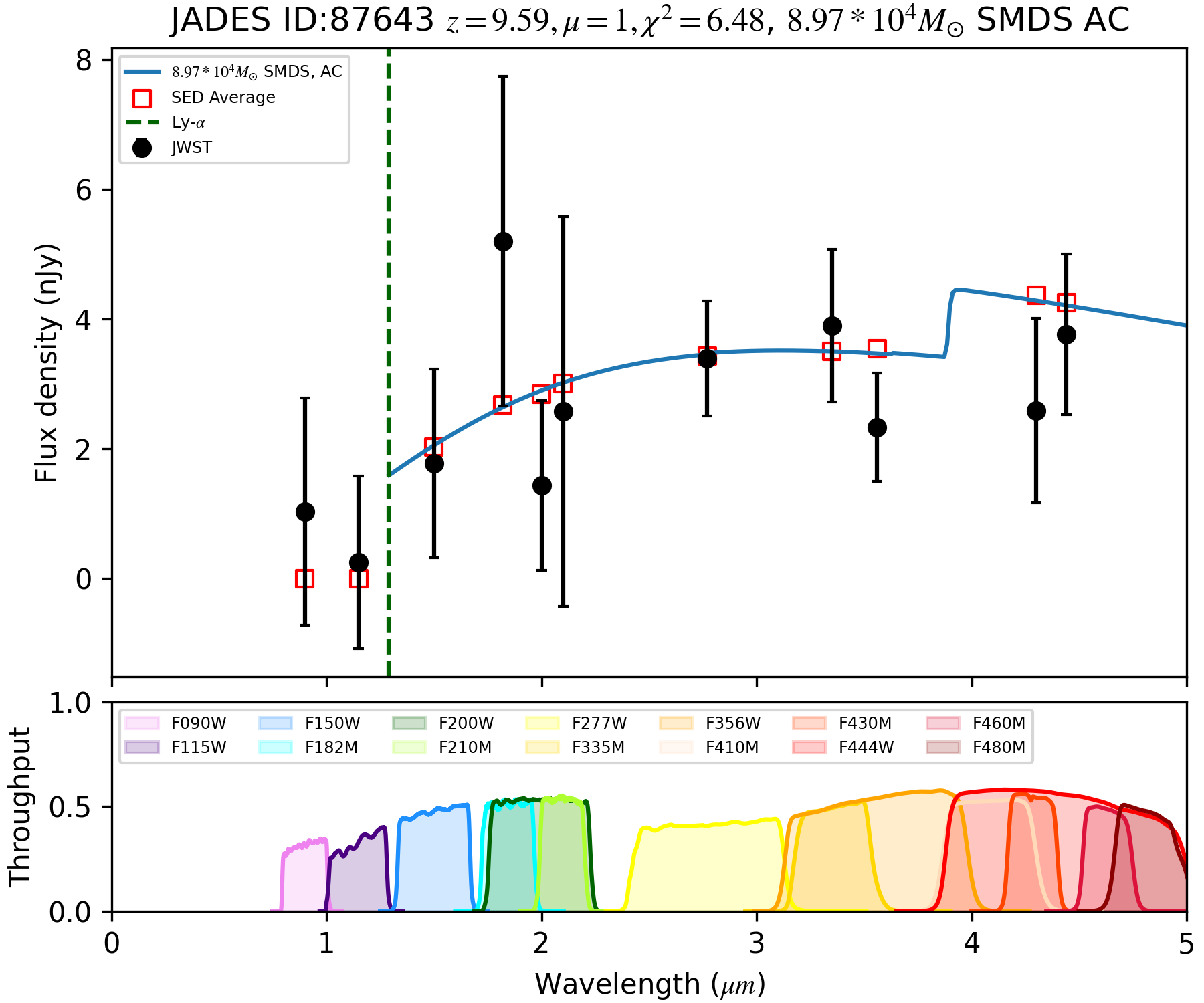}\\[-0.5ex]{\scriptsize (c)}
\end{minipage}\\[0.3ex]
\begin{minipage}{0.32\textwidth}\centering
\includegraphics[width=\linewidth]{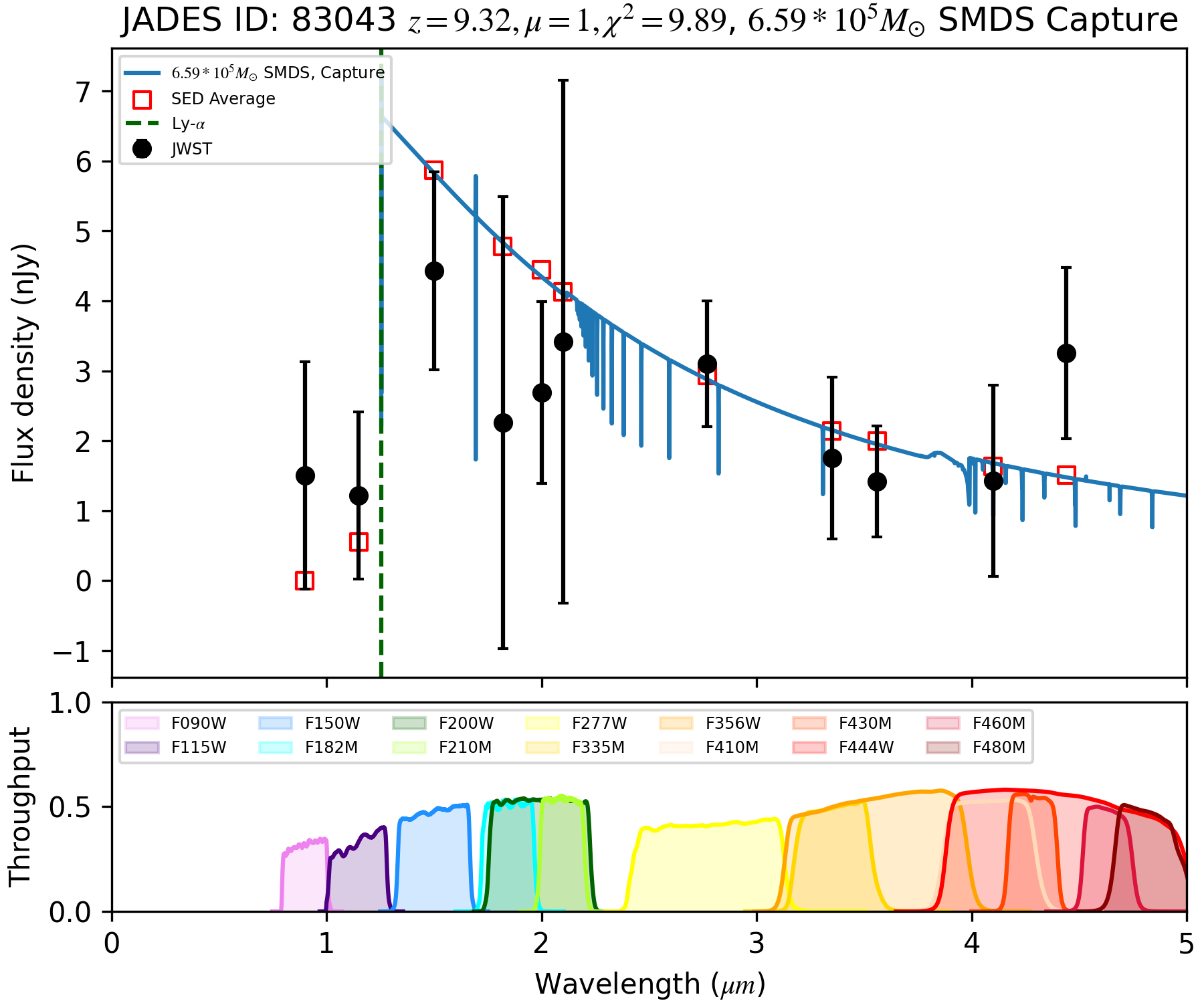}\\[-0.5ex]{\scriptsize (d)}
\end{minipage}\hfill
\begin{minipage}{0.32\textwidth}\centering
\includegraphics[width=\linewidth]{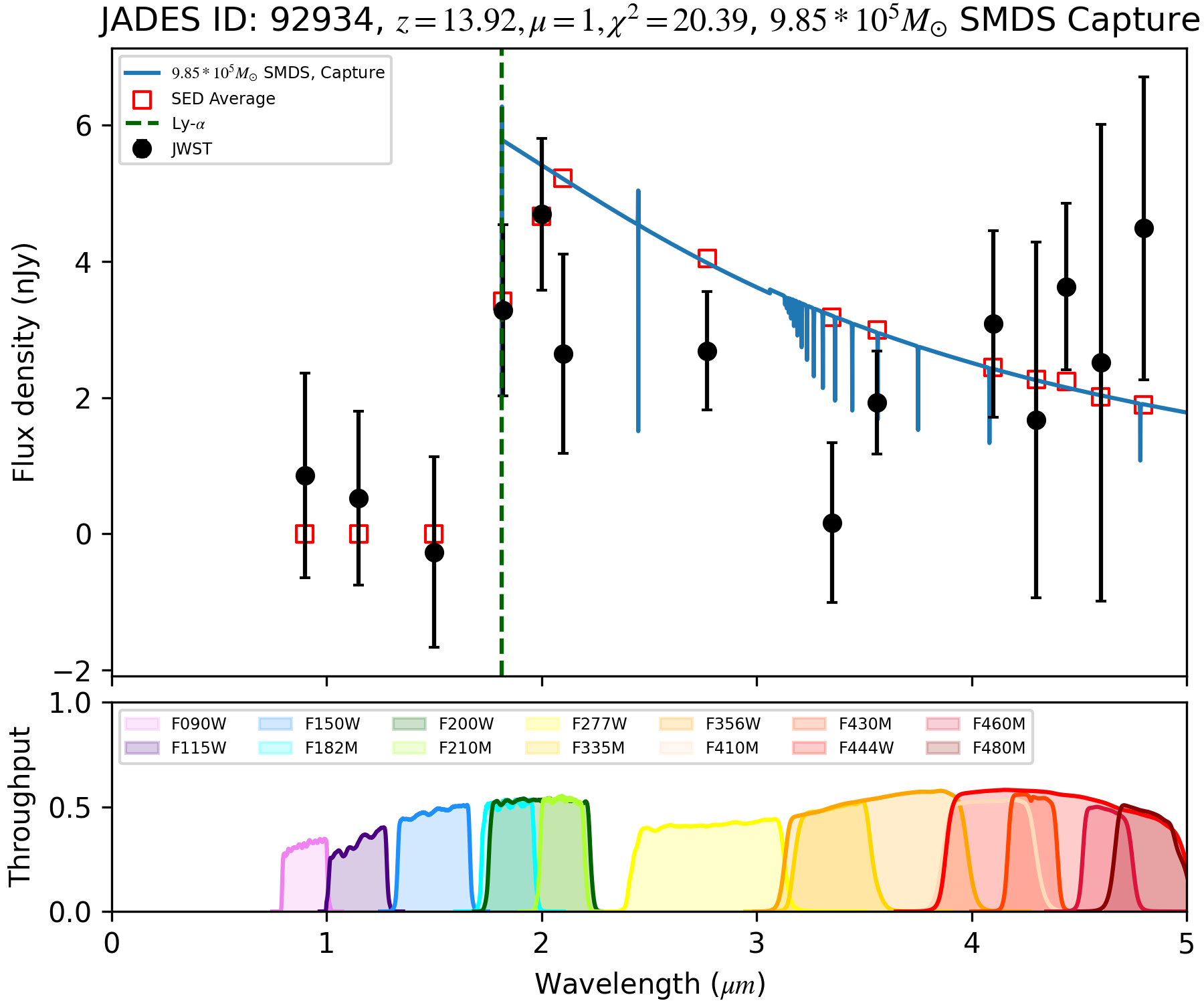}\\[-0.5ex]{\scriptsize (e)}
\end{minipage}\hfill
\begin{minipage}{0.32\textwidth}\centering
\includegraphics[width=\linewidth]{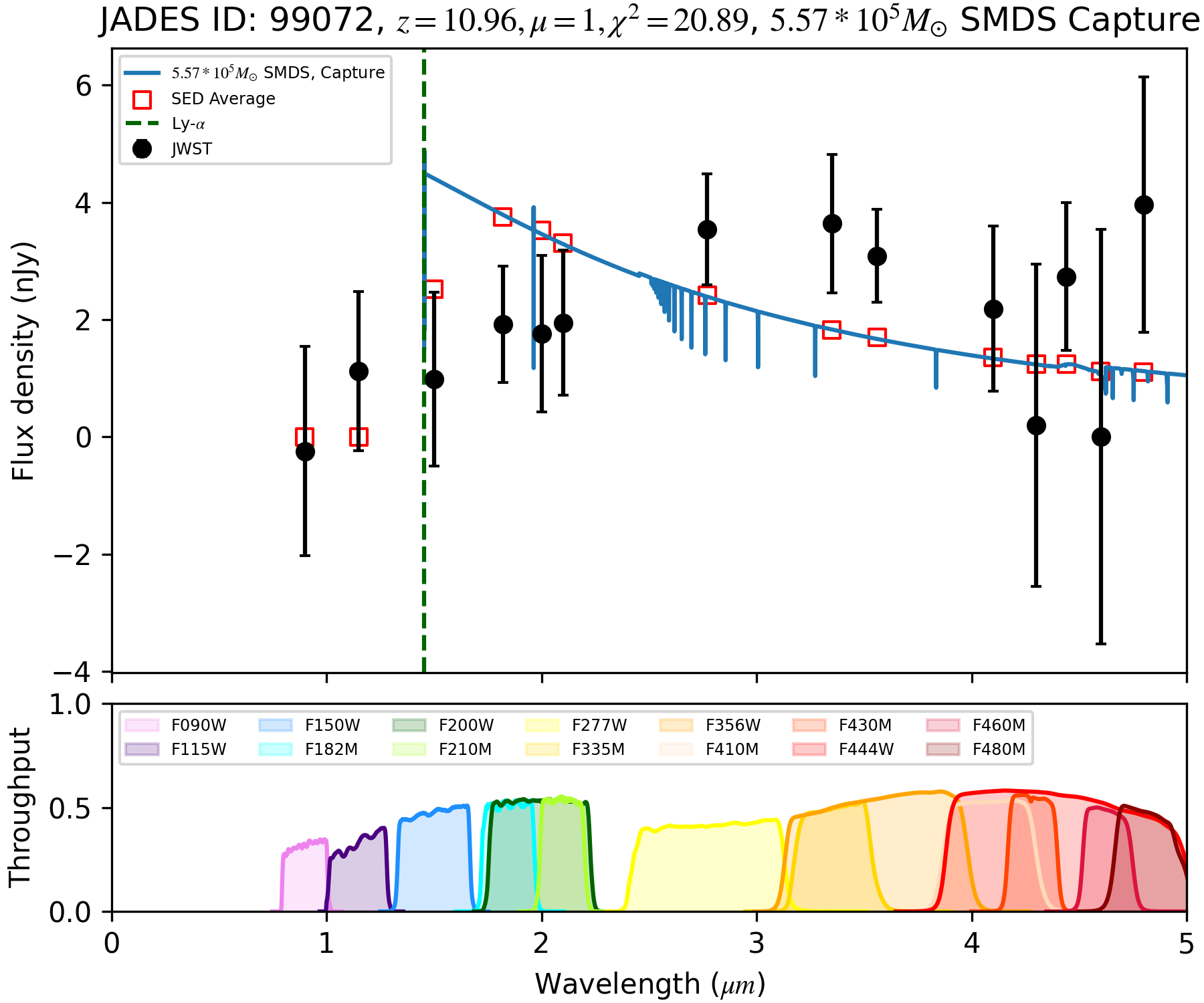}\\[-0.5ex]{\scriptsize (f)}
\end{minipage}
\caption{The six new photometric SMDSs candidates identified in this work. JWST data (black dots with error bars) plotted against our FFNN predicted SMDS SEDs (bue lines). Top row, i.e. panels a-c, depict three high-z JADEs objects that have photometry consistent with SMDSs formed via the extended AC mechanism. Bottom row, i.e. panels d-f, present three JADES objects that have photometry consistent with SMDSs powered by captured DM. On the title of each plot we summarize: object ID, photometric redshift (z) found by our FFNN, value of the gravitational boost factor ($\mu$), $\chi^2$ statistic, SMDSs mass in $\Msun$, and type of SMDS model (AC or Capture).\label{fig:newcandidates}}
\end{figure*}

\begin{table*}[!tbp]
    \centering
    \caption{Summary of fit parameters for the six new photometric candidates found in this work. Reported are the inferred masses (in $10^5,M_\odot$), formation mechanism for the SMDS (AC or DM Capture), best fit redshifts, $\chi^2$ values, and the corresponding 95\% $\chi^2_{\rm critical}$ thresholds.}
    \label{tab:comparison}
    \begin{tabular}{lcccccr}
        \hline
        \textbf{Object Name} & \textbf{Mass ($M_\odot$)}& \textbf{SMDS(AC/Capture)} &\textbf{$z_{phot}$}  & \texttt{\#} of NIRCam Bands &\textbf{$\chi^2$} & \textbf{$\chi_{critical}^2$}\\
        \hline
        JADES ID 93735 & $ 5.2\times 10^4$ &Adiabatic Contraction&8.24 & 11 &15.88 & 18.3 \\
        JADES ID 94961 & $1.15 \times 10^5$ &Adiabatic Contraction &8.92 & 14 &19.95 & 22.4 \\
        JADES ID 87643 & $8.97\times 10^4$ &Adiabatic Contraction &9.59 & 11 &6.48 & 18.3 \\
        JADES ID 83043 & $6.59 \times 10^5$ & DM Capture&9.32 & 11 &9.89 & 18.3 \\
        JADES ID 92934 & $9.85 \times 10^5$ & DM Capture&13.92 & 14 &20.39 &22.4 \\
        JADES ID 99072 & $5.57 \times 10^5$ & DM Capture&10.96 & 14 &20.89 &22.4 \\
         \hline
    \end{tabular}
\end{table*}

Fig.~\ref{fig:newcandidates} presents a detailed comparison of the predicted and observed SEDs for adiabatic contraction (panels a-c) and DM capture (panels d-f) photometric candidates newly identified in this work. The solid lines represent the neural network’s predictions, while the black points with error bars denote JWST observations. For all candidates, we plot the dark star SEDs for the given mass and redshift that captures both the overall trends and finer details across wavelengths. These results highlight the neural network's ability to infer key physical parameters such as stellar mass and redshift, even in the presence of observational uncertainties. A summary of the relevant parameters for the six new photometric candidates identified is presented in Table~\ref{tab:comparison}. 

Our findings not only demonstrate the effectiveness of the neural network in identifying dark star candidates but also underscore its computational efficiency compared to traditional methods. The model’s ability to handle incomplete datasets without compromising accuracy offers a significant advantage for large-scale photometric surveys. By streamlining the detection process, we open new avenues for identifying dark star photometric candidates suitable for spectroscopic followup.

\section{Limitations} \label{sec: Limitations}

While the neural network presented in this study demonstrates remarkable accuracy in identifying Dark Star candidates and predicting their physical properties, several opportunities exist to further enhance its capabilities.
One significant area for improvement is the incorporation of observational uncertainties. Dark Star candidates, observed at very high redshifts, are often associated with photometric data that include substantial uncertainties. The current architecture does not account for these error bounds, which may affect the precision of the mass and redshift estimates. In future work, we aim to address this limitation by developing a Bayesian Neural Network (BNN) \citep{mackay1992bnn, gal2016dropout}. Unlike standard neural networks, BNNs incorporate uncertainties directly into their analysis, offering both point estimates and credible intervals for predictions. This enhancement will allow for more robust and scientifically grounded predictions, further strengthening the reliability of the model. Moreover, we plan to include in an updated version of the FFNN model developed here the possibility to analyze photometric data from the MIRI instrument onboard JWST, further increasing its constraining power.

We also envision automating the process of identifying Dark Star candidates. To achieve this, we aim to integrate the neural network directly with the JADES catalog, enabling the automated identification of potential candidates. Such a pipeline would streamline the workflow, maximize time efficiency, and allow for the real-time analysis of vast datasets, making it a powerful tool for future astronomical surveys.

Moreover, the current model focuses primarily on predicting mass and redshift, leaving out other critical physical characteristics of Dark Stars. Features such as nebular emission, which may contribute to their observational signatures (such as flux and size), will be incorporated into future iterations of the model. Moreover, we plan to model SMDSs powered by WIMPs of other masses rather the commonly assumed and used 100 GeV in the Dark Stars literature. In order to achieve this we would first need to model SMDSs powered by WIMPs of various masses with \texttt{MESA}, an objective we are currently working on. By expanding the input dataset to include these additional possible effects and retraining the neural network on this enriched parameter space, we aim to significantly improve its predictive power and applicability to a broader range of astrophysical scenarios. This holistic approach will enable a more comprehensive study of Dark Stars, enhancing both the scope and depth of the model's scientific insights.

\section{Conclusions} \label{sec: Conclusions}

This study demonstrates the efficacy of neural networks in identifying Dark Star candidates and predicting their physical properties, such as stellar mass and redshift, using photometric data from the JADES survey. By employing a robust architecture capable of handling both complete and incomplete datasets, we successfully identified promising candidates for both adiabatic contraction and capture case formation mechanisms. The results show excellent agreement between the predicted and observed flux densities, with $\chi^2$ values falling within the 95\% confidence interval, affirming the reliability of the model.

With photometry alone it is unlikely one can disambiguate between SMDSs and high-z galaxies. The only known way tell those two kind of sources apart is using spectroscopy. For example, early galaxies are expected to contain a large number of hot stars, and as such, strong nebular emission lines, such as the Lyman-alpha emission. In contrast, Dark Stars, even those powering a nebula, would not have such a Lyman-alpha emission line. Moreover, isolated dark stars are zero metallicity systems, so nothing besides H and He lines should be present in their spectra. Conversely, early galaxies can exhibit strong metal emission lines such as [O~III] lines which can be targeted with ALMA.  The only way to confirm a SDMS is via the observation of its smoking gun absorption feature at 1640 \AA (restframe), which is due to  large amounts of singly ionized Helium (He~II) in its first excited state in its atmosphere. Moreover, spectral data (even when noisy) is more constraining than photometry. With this in mind, in the companion paper~\citep{NNSMDSSpectra}, we develop a neural network adapted specifically for JWST NIRSpec data. 

The ability to detect and analyze Dark Stars at high redshifts opens new avenues for understanding the early universe and the role of these exotic objects in cosmic evolution. This work represents an important step toward the integration of machine learning into astrophysical research, specifically, for the applying for the first time AI NN methods to the detection of Dark Star candidates. With continued advancements, neural networks hold the potential to revolutionize the search for and study of Dark Stars, paving the way for exciting discoveries in the future.

\section{Data Availability}
The data used in this work include products from the JWST Advanced Deep Extragalactic Survey (JADES) public release hosted by MAST \citep{MAST_JADES_10.17909_8tdj_8n28}.

\begin{acknowledgments}
 C.I. acknowledges funding from Colgate University
via the Research Council (Grant No. 821028) and the Picker Interdisciplinary Science Institute (Grant No. 826837).  We furthermore acknowledge the use of Colgate’s Turing Supercomputer 
(Partially supported by NSF grant OAC-2346664).
\end{acknowledgments}

\bibliography{sample631,RefsDM}{}
\bibliographystyle{elsarticle-harv}



\end{document}